\documentclass{aa}
\usepackage{graphicx,amsmath}
\usepackage{epsf}
\usepackage{txfonts}
\usepackage{geometry}
\usepackage{tablefootnote}
\usepackage{longtable}
\usepackage{lscape}
\usepackage[hyperindex,breaklinks=true, colorlinks, citecolor=blue]{hyperref}  
\usepackage{comment}
\usepackage{natbib}
\usepackage{upgreek}
\usepackage{float}
\usepackage[switch,pagewise]{lineno}
\usepackage{multirow}

\newcommand{\CO}{$^{12}$CO(1--0)}
\newcommand{\kps}{km\,s$^{-1}$}
\providecommand{\sorthelp}[1]{}

\def\deg{\ifmmode^\circ\else$^\circ$\fi}
\def\pdeg{\ifmmode $\setbox0=\hbox{$^{\circ}$}\rlap{\hskip.11\wd0 .}$^{\circ}
          \else \setbox0=\hbox{$^{\circ}$}\rlap{\hskip.11\wd0 .}$^{\circ}$\fi}
\def\arcs{\ifmmode {^{\scriptstyle\prime\prime}}
          \else $^{\scriptstyle\prime\prime}$\fi}
\def\arcm{\ifmmode {^{\scriptstyle\prime}}
          \else $^{\scriptstyle\prime}$\fi}
\newdimen\sa  \newdimen\sb
\def\parcs{\sa=.07em \sb=.03em
     \ifmmode \hbox{\rlap{.}}^{\scriptstyle\prime\kern -\sb\prime}\hbox{\kern -\sa}
     \else \rlap{.}$^{\scriptstyle\prime\kern -\sb\prime}$\kern -\sa\fi}
\def\parcm{\sa=.08em \sb=.03em
     \ifmmode \hbox{\rlap{.}\kern\sa}^{\scriptstyle\prime}\hbox{\kern-\sb}
     \else \rlap{.}\kern\sa$^{\scriptstyle\prime}$\kern-\sb\fi}

\begin{document}

\title{Anisotropy in the carbon monoxide (CO) line emission across the Milky Way's disk}
\titlerunning{CO filament orientation}
    \author{
        J.~D.~Soler$^{1}$\thanks{Corresponding author, \email{juandiegosolerp@gmail.com}},
        M.~Heyer$^{2}$,
        M.~Benedettini$^{1}$
        D.~Elia$^{1}$,
        P.~Hennebelle$^{3}$
        R.~S.~Klessen$^{4,5,6,7}$,
        C.~Mininni$^{1}$,
        A.~Nucara$^{1,8}$,
        V.-M.~Pelkonen$^{1}$,
        S.~Molinari$^{1}$,
        R. J. Smith$^{9}$,
        E.~Schisano$^{1}$,
        A.~Traficante$^{1}$,
        R.~Tre{\ss}$^{10}$
} 
\institute{
1. Istituto di Astrofisica e Planetologia Spaziali (IAPS). INAF. Via Fosso del Cavaliere 100, 00133 Roma, Italy\\
2. Department of Astronomy, University of Massachusetts, Amherst, MA 01003, USA\\
3. Laboratoire AIM, Paris-Saclay, CEA/IRFU/SAp - CNRS - Universit\'{e} Paris Diderot. 91191, Gif-sur-Yvette Cedex, France\\
4. Universit\"{a}t Heidelberg, Zentrum f\"{u}r Astronomie, Institut f\"{u}r Theoretische Astrophysik, Albert-Ueberle-Str. 2, 69120, Heidelberg, Germany\\ 
5. Universit\"{a}t Heidelberg, Interdiszipli\"{a}res Zentrum f\"{u}r Wissenschaftliches Rechnen, Im Neuenheimer Feld 225, 69120 Heidelberg, Germany\\
6. Harvard-Smithsonian Center for Astrophysics, 60 Garden Street, Cambridge, MA 02138, USA\\
7. Elizabeth S. and Richard M. Cashin Fellow at the Radcliffe Institute for Advanced Studies at Harvard University, 10 Garden Street, Cambridge, MA 02138, USA\\
8. Dipartimento di Fisica, Università di Roma Tor Vergata, Via della Ricerca Scientifica 1, I-00133 Roma, Italy\\
9. SUPA, School of Physics and Astronomy, University of St Andrews, North Haugh, St Andrews, KY16 9SS, UK \\
10. \'{E}cole Polytechnique F\'{e}d\'{e}rale de Lausanne, Observatoire de Sauverny, Chemin Pegasi 51, CH-1290 Versoix, Switzerland
}
\authorrunning{Soler,\,J.D. et al.}

\date{Received XX XX XXXX / Accepted XX XX XXXX}

\abstract{
We present a study of the CO line emission anisotropy across the Milky Way's disk to examine the effect of stellar feedback and Galactic dynamics on the distribution of the dense interstellar medium.
The Hessian matrix method is used to characterize the \CO\ line emission distribution and identify the preferential orientation across line-of-sight velocity channels in the Dame et al. 2001 composite Galactic plane survey, which covers the Galactic latitude range $|b|$\,$<$\,5\deg.
The structures sampled with this tracer are predominantly parallel to the Galactic plane toward the inner Galaxy, in clear contrast with the predominantly perpendicular orientation of the structures traced by neutral atomic hydrogen (H{\sc i}) emission toward the same regions.
The analysis of the Galactic plane portions sampled at higher angular resolution with other surveys reveals that the alignment with the Galactic plane is also prevalent at smaller scales.
We find no preferential orientation in the CO emission toward the outer Galaxy, in contrast with the preferential alignment with the Galactic plane displayed by H{\sc i} in that portion of the Milky Way.
We interpret these results as the combined effect of the decrease in mid-plane pressure with increasing Galactocentric radius and SN feedback lifting diffuse gas more efficiently than dense gas off the Galactic plane.
}

\maketitle

\section{Introduction}

Carbon monoxide (CO) emission is a crucial tracer of the cold, dense gas related to star formation \citep{kennicuttANDevans2012,heyerANDdame2015}.
Molecular clouds (MCs) traced by CO have been broadly sampled in the Milky Way and nearby galaxies \citep[see, for example,][]{fukui2010,heyerANDdame2015,schinnerer2024}.
Still, many aspects of the MC lifecycle remain uncertain.
For example, the impact of stellar winds, radiation, and supernovae (SNe) interacting with the surrounding gas on different scales and the role this interaction plays in the formation and dissipation of MCs are not established \citep[see, for example,][]{elmegreen2000,krumholz2014,klessen2016,maclow2017}.
In this paper, we investigate the Galactic processes influencing the MC lifecycle by studying the anisotropy in the CO line emission distribution throughout the Galactic plane, extending to the molecular gas phase the study of the neutral atomic hydrogen (H{\sc i}) emission presented in \citet{soler2022}.

High star formation rate (SFR) densities can produce enough energy and momentum to launch outflows of ionized, neutral, and molecular gas that can potentially escape the main body of a galaxy \citep[see, for example,][]{dekelANDsilk1986,krumholz2017}.
These galactic outflows remove the raw material for future star formation and enrich the galactic disk and circumgalactic medium with heavy metals, thus driving the evolution of galaxies \citep[see, for example,][]{veilleux2005,fabian2012,saintonge2022}.
Starburst-driven molecular winds have been observed in emission stacks of nearby ultra-luminous infrared galaxies \citep[see, for example,][]{chung2011}. 
They are resolved in near the disks of local starburst galaxies, such as NGC253 \citep[$d$\,$=$\,3.5\,Mpc;][]{sakamoto2006,bolatto2013,krieger2019}, M82 \citep[$d$\,$=$\,3.9\,Mpc;][]{weiss1999,walter2002,matsushita2005,leroy2015}, NGC1808 \citep[$d$\,$=$\,10.8\,Mpc;][]{salak2018}, NGC2146 \citep[$d$\,$=$\,17.2\,Mpc;][]{tsai2009,kreckel2014}, NGC3256 \citep[$d$\,$=$\,35\,Mpc;][]{sakamoto2006}, and ESO320-G030 \citep[$d$\,$=$\,48\,Mpc;][]{pereira-santaella2016}.
\cite{stuber2021} reported that almost a quarter of the 90 nearby main-sequence galaxies covered in the Physics at High Angular resolution in Nearby GalaxieS (PHANGS) ALMA Survey show evidence of central molecular winds.

The mean SFR in these starburst galaxies exceeds that of the Milky Way by at least two orders of magnitude \citep[][]{elia2022,zari2023,soler2023b,elia2025}.
Yet, the Galactic SFR may induce weaker outflows that shape the molecular gas and drive metallicity changes throughout the Galactic disk \citep[see, for example,][]{haywardANDhopkins2017,kimANDostriker2018,sharda2024}.
Studies of blown-out molecular structures in the Milky Way have been limited to the central region of the Galaxy \citep[see, for example, ][]{diteodoro2020,heyer2025}.
A systematic study of their prevalence throughout the disk is not yet available.

\cite{soler2022} used a morphological decomposition in terms of filaments to study the anisotropy in the H{\sc i} emission toward the Milky Way's disk using the 16\parcm2-resolution observations in the H{\sc i} 4$\pi$ (HI4PI) survey \citep{hi4pi2016}.
That work identified a transition in the preferential orientation of the H{\sc i} filamentary structures with Galactocentric radius, from mostly perpendicular or no preferred orientation with respect to the Galactic plane for Galactocentric radii $R_{\rm gal}$\,$<$\,10\,kpc to mostly parallel for $R_{\rm gal}$\,$>$\,10\,kpc.
By comparison with the populations of high-mass stars and identified SN bubbles, as well as the results obtained in numerical magnetohydrodynamic (MHD) simulations of a multiphase stratified Galactic medium, the authors interpreted this result as the imprint of SN feedback in the inner Galaxy and Galactic rotation and shear in the outer Milky Way \citep{soler2020}.
In this paper, we apply the method introduced in \cite{soler2022} to study the anisotropy in the distribution of CO emission across the Galactic plane and identify the prevalence of vertical molecular structures that may indicate molecular galactic outflows.

Extended surveys of the CO emission toward the Galactic plane enabled the discovery of long, high-density filamentary features that might be shaped by the structural dynamics of the Milky Way, for example, the Nessie cloud \citep{goodman2014}.
These objects, usually associated with filamentary infrared dark clouds (IRDCs), have been identified and cataloged using a variety of methods \citep[see, for example,][]{ragan2014,abreu-vicente2016,wang2016}.
Their lengths range between tens and a few hundred parsecs, masses between $10^{3}$ and $10^{6}$\,M$_{\odot}$, aspect ratios that range between 3:1 and 117:1, and are preferentially oriented parallel to the Galactic plane \citep{zucker2018}.
More recently, \cite{neralwar2022a} and \cite{neralwar2022b} studied the morphology of clouds in the $^{12}$CO(2--1) observations in the Structure, Excitation and Dynamics of the Inner Galactic Interstellar Medium (SEDIGISM) survey \citep{schuller2021} and found that most of them are elongated, according to the J plots classification algorithm \citep{jaffa2018}.

This work focuses on the preferential orientation of filamentary structures in the CO emission to identify trends across LOS velocities ($v_{\rm LSR}$).
Thus, we do not consider the physical properties of individual filamentary structures, recently reviewed in \cite{hacar2023}, but rather the orientation of high-aspect-ratio features in the emission as a marker of anisotropy in the velocity field.
Such a study has never been performed in CO across the Galactic plane, employing circular statistics tools to quantify the emission anisotropy.

This paper is organized as follows.
Section~\ref{section:observations} presents the CO observations used in this study.
In Sec.~\ref{section:methods}, we introduce the Hessian matrix method for identifying the filamentary structures and the circular statistics employed to quantify its anisotropy.
Sec.~\ref{section:results} describes the global trends in CO anisotropy and its relation with the H{\sc i} emission.
We discuss the implications of these results for our understanding of the Galactic structure in Sec.~\ref{section:discussion}.
Finally, in Sec.~\ref{section:conclusions}, we present our conclusions. 
We complement the main results of this work with the analysis shown in a set of appendices.
Appendix~\ref{appendix:method} presents details on the HOG method's error propagation and selection of parameters.
In App.~\ref{app:gradients}, we show that the reported filament anisotropy in the CO emission can also be recovered using the preferential orientation of intensity gradients.
We present a study of synthetic emission maps produced using fractional Brownian motion (fBm) realizations in App.~\ref{app:fBM}, which we use to illustrate the effects of the beam size in the Hessian analysis results.

\begin{figure*}[ht]
\centerline{\includegraphics[width=0.995\textwidth,angle=0,origin=c]{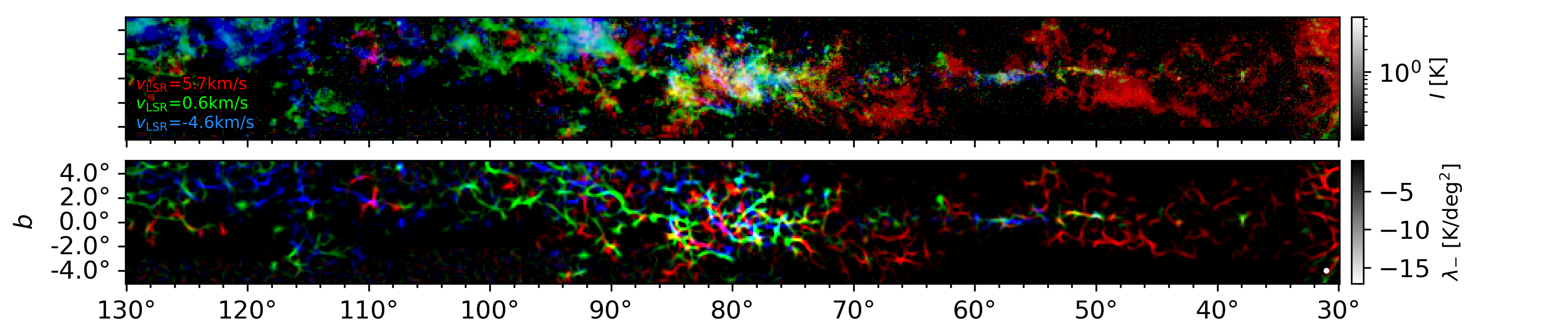}}
\caption{Example of the filamentary structure identification in the CO emission toward a portion of the Galactic plane.}
\label{fig:exampleTile}
\end{figure*}

\section{Data}\label{section:observations}

CO emission surveys do not uniformly cover the Galactic plane \citep[see, for example,][]{heyerANDdame2015}.
The best angular-resolution survey of the CO line emission covering the whole Galactic plane with the same isotopologue and at the same transition is in the \CO\ compilation presented in \cite{dame2001}.
That is the core dataset that we used in this analysis.

We also considered other surveys to extend our study to higher (sub-arcminute) angular resolutions for selected sections of the Galactic plane.
First, the Milky Way Imaging Scroll Painting (MWISP) survey, which sampled the \CO, $^{13}$CO($J$=1--0), and C$^{18}$O ($J$=1--0) line emission toward the Galactic and has so far released its observations for the region within $|b|$\,$<$\,1\deg for  25\pdeg0\,$<$\,$l$\,$<$\,49\pdeg0
Second, the Forgotten Quadrant Survey \citep[FQS;][]{benedettini2020}, which sampled the \CO\ line emission in the range 220\deg\,$<$\,$l$\,$<$\,240\deg\ and $-2\pdeg5$\,$<$\,$b$\,$<$\,0\deg.

We acknowledge other higher-resolution \CO\ surveys covering different portions of the plane, for example, the FOREST Unbiased Galactic plane Imaging survey with the Nobeyama 45-m telescope \citep[FUGIN;][]{unemoto2017} and the Census of High- and Medium-mass Protostars (CHaMP) survey \citep{barnes2016}.
However, our goal is to illustrate examples of the reported \CO\ anisotropies at higher resolutions rather than produce a global anisotropy analysis at higher resolutions, which requires the combination of a dissimilar variety of datasets beyond this work's scope.

\subsection{The Dame et al. (2001) composite CO survey of the Milky Way disk.}

The \CO\ line emission observations presented in \cite{dame2001} encompass the surveys obtained over two decades with the Harvard-Smithsonian Center for Astrophysics (CfA)'s 1.2-m telescope and a similar telescope on Cerro Tololo in Chile.
These observations have an angular resolution of 8\parcm5 at 115\,GHz, the frequency of the \CO\ line.

For this study, we used the interpolated composite Galactic plane survey, which encompasses the range 0.0\,$<$\,$l$\,$<$\,360.0\deg\ and $|b|$\,$\leq$\,5\deg\ with 1.3-km\,s$^{-1}$-wide spectral channels.
The noise level throughout this dataset is not uniform, as it combines surveys with different instruments and acquired at various times.
The common noise level estimation for this data set was discussed in appendix~A of \cite{miville-deschenes2017}, where the authors identified three peaks in the noise distribution at 0.06, 0.10, and 0.19\,K per channel.
We adopt the latter as the global noise level for this dataset.

We used the {\tt astropy} {\tt reproject} package to project this data into a spatial grid covering the range 0\pdeg0\,$<$\,$l$\,$<$\,360\pdeg0 and $|b|$\,$<$\,$5\pdeg0$ with a pixel size $\Delta l$\,$=$\,$\Delta b$\,=7\parcm5.
We also used the {\tt astropy} {\tt spectral-cube} package to project the spectral axis of these observations into the 1.29-km\,s$^{-1}$-resolution spectral grid of the H{\sc i}4PI survey \citep[][]{hi4pi2016}.
This spectral reprojection allows the direct comparison with the results presented in \cite{soler2022} and does not significantly alter the results of our analysis.

\subsection{The Milky Way Imaging Scroll Painting (MWISP) survey}

The MWISP project is a high-sensitivity survey of the northern Galactic plane observed with the Purple Mountain Observatory 13.7-m telescope \citep{sun2018,su2019}.
It comprises the $^{12}$CO(1--0), $^{13}$CO(1--0), and C$^{18}$O (1--0) lines simultaneously observed by the nine-beam Superconducting Spectroscopic ARray (SSAR) receiving system \citep{shan2012}.
The full-width half maximum (FWHM) of the observations is 49\arcs\ at the $^{12}$CO frequency and 51\arcs\ at the $^{13}$CO and C$^{18}$O frequencies, respectively.

The SSAR bandwidth of 1\,GHz with 16,384 channels provides a velocity coverage of 260\,\kps\ and a spectral resolution of 61\,kHz, equivalent to velocity separations of about 0.16\,\kps\ for $^{12}$CO and 0.17\,\kps\ for $^{13}$CO and C$^{18}$O.
This velocity range enables the sampling of $v_{\rm LSR}$\,$<$\,0\,\kps\ towards the first Galactic quadrant (QI), which is excluded in other high-resolution CO surveys of the Galactic plane, such as the Boston University-Five College Radio Astronomy Observatory Galactic Ring Survey \citep[GRS;][]{jackson2006} and FUGIN \citep{unemoto2017}.
The typical MWISP root-mean-square (RMS) noise levels are about 0.5\,K for $^{12}$CO and 0.3\,K for $^{13}$CO and C$^{18}$O, respectively.
The final data products are position-position-velocity (PPV) cubes constructed from a mosaic with a spatial grid spacing of 30\arcsec.

\subsection{The Forgotten Quadrant Survey (FQS)}

The FQS is the product of 700 hours of observations with the Arizona Radio Observatory (ARO) 12-m antenna.
It covered the Galactic plane in the range 220\deg\,$<$\,$l$\,$<$\,240\deg and $-$2\pdeg5\,$<$\,$b$\,$<$\,0\deg sampling the \CO\ and $^{13}$CO(1--0) emission.
The survey was divided into partially overlapping 30\arcmin\,$\times$\,30\arcmin\ tiles with sides aligned along the Galactic longitude and latitude.
Each tile was observed twice in the On-The-Fly observing mode in mutually orthogonal scan directions, one along $l$ and one along $b$.
The angular resolution of these observations is set by the 55\arcsec\ telescope beam at 115\,GHz.
The data was acquired by scanning rows separated by 18\arcsec.
The FQS \CO\ data product is distributed in a grid with 17\parcs3 pixels and 0.65-\kps-wide velocity channels.
Its RMS noise level ranges from 0.8\,K to 1.3\,K per channel.
Further data acquisition and reduction information is presented in \cite{benedettini2020}.

\begin{figure*}[ht]
\centerline{
\includegraphics[width=0.495\textwidth,angle=0,origin=c]{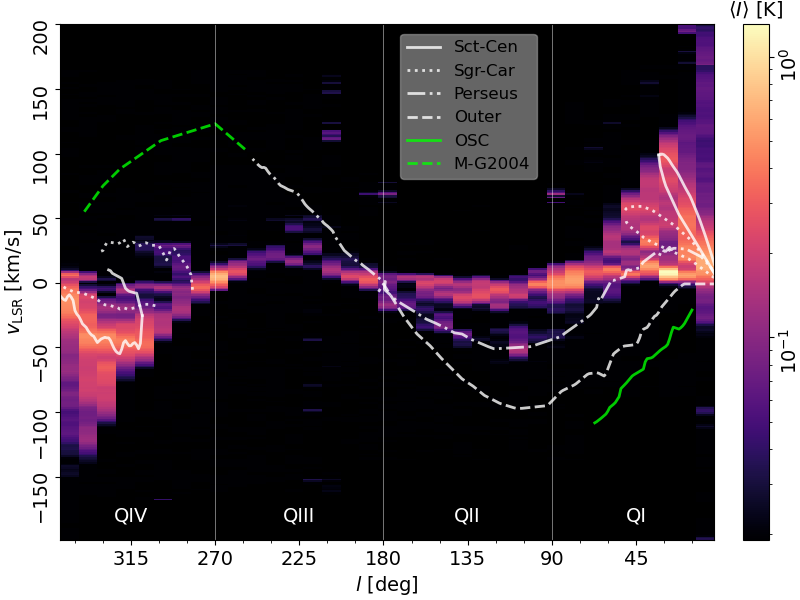}
\includegraphics[width=0.495\textwidth,angle=0,origin=c]{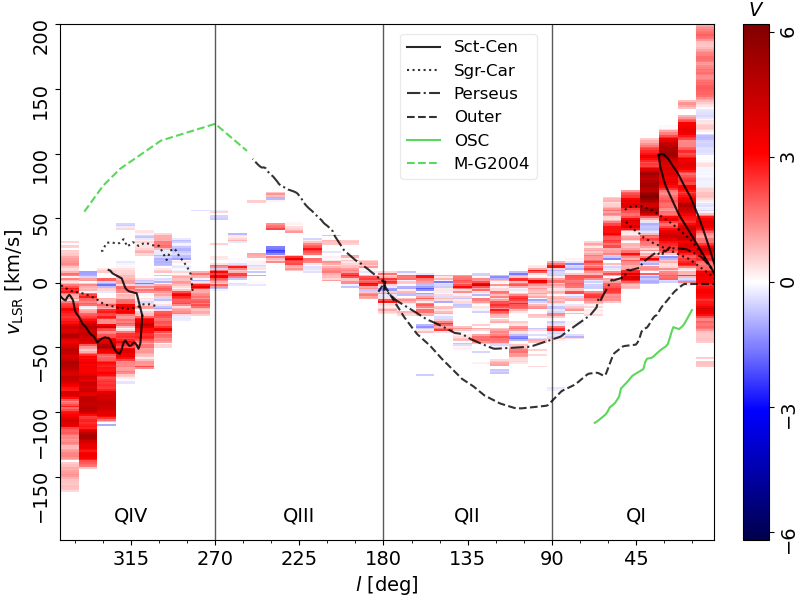}
}
\caption{Longitude-velocity ($lv$) diagrams of the CO mean intensity ($\left<I\right>$, left) and filament orientation anisotropy quantified by the projected Rayleigh statistic ($V$, right).
Each of the pixel elements in the diagrams corresponds to a 10\deg\,$\times$\,10\deg\,$\times$\,1.29\,km\,s$^{-1}$ velocity channel map centered on $b$\,$=$\,0\deg, which we call tile throughout out this paper.
These results correspond to the Hessian analysis using an 18\arcmin\ FHWM derivative kernel.
Values of $V$\,$>$\,0 (red) or $V$\,$<$\,$0$ (blue) indicate a preferential orientation of the filaments parallel, $\theta$\,$=$\,$0$\deg, or perpendicular, $\theta$\,$=$\,$90$\deg, to the Galactic plane.
The 3$\sigma$ statistical significance for these two orientations corresponds to $V$\,$>$\,2.87 or $V$\,$<$\,$-2.87$, respectively.
The overlaid curves correspond to the main spiral arms features presented in \cite{reid2016}, the outer Scutum-Centaurus arm \citep[OSC,][]{dameANDthaddeus2011}, and the extended outer arm \citep[M-G2004,][]{mcclure-griffiths2004}.
}
\label{fig:lvdiagrams1}
\end{figure*}

\section{Methods}\label{section:methods}

We applied the Hessian matrix method for filament identification as follows.
We considered the CO line emission maps across velocity channels $I_{\rm CO}(l,b,v_{\rm LSR})$, where the position in the sky is identified by the Galactic longitude and latitude, $l$ and $b$, and $v_{\rm LSR}$ is the line-of-sight velocity with respect to the local standard of rest (LSR).
For each $v_{\rm LSR}$ channel, we estimated the derivatives with respect to the local coordinates $(x,y)$ and built the Hessian matrix,
\begin{equation}
\mathbf{H}(x,y) \equiv \begin{bmatrix} 
H_{xx} & H_{xy} \\
H_{yx} & H_{yy} 
\end{bmatrix},
\end{equation}
where $H_{xx}$\,$\equiv$\,$\partial^{2} I/\partial x^{2}$, $H_{xy}$\,$\equiv$\,$\partial^{2} I/\partial x \partial y$, $H_{yx}$\,$\equiv$\,$\partial^{2} I/\partial y \partial x$, $H_{yy}$\,$\equiv$\,$\partial^{2} I/\partial y^{2}$, and $x$ and $y$ are related to the Galactic coordinates $(l, b)$ as $x$\,$\equiv$\,$l\cos b$ and $y$\,$\equiv$\,$b$, so that the $x$-axis is parallel to the Galactic plane.
In the sky areas considered in this study, $|b|$\,$\leq$\,5\deg, $\cos b$\,$\approx$\,1, so the derivatives are performed on a tangent plane projection of each tile, where $x$\,$\equiv$\,$l$ and $y$\,$\equiv$\,$b$.

We obtained the partial spatial derivatives using Gaussian derivatives, that is, by convolving $I(l,b,v)$ with the second derivatives of a two-dimensional Gaussian function, following the procedure described in \cite{soler2013}.
To match the angular scales considered in \cite{soler2022}, we considered an 18\arcmin\ FWHM derivative kernel.
The results obtained with different derivative kernel sizes are presented in Appendix~\ref{appendix:method}.

The two eigenvalues ($\lambda_{\pm}$) of the Hessian matrix were found by solving the characteristic equation,
\begin{equation}\label{eq:lambda}
\lambda_{\pm} = \frac{(H_{xx}+H_{yy}) \pm \sqrt{(H_{xx}-H_{yy})^{2}+4H_{xy}H_{yx}}}{2}.
\end{equation}
Both eigenvalues define the local curvature of the intensity map.
In particular, the minimum eigenvalue ($\lambda_{-}$) highlights filamentary structures or ridges, as detailed in \cite{planck2014-XXXII}.
The eigenvector corresponding to $\lambda_{-}$ defines the orientation of intensity ridges with respect to the Galactic plane, which is characterized by the angle
\begin{equation}\label{eq:theta}
\theta = \frac{1}{2}\arctan\left[\frac{H_{xy}+H_{yx}}{H_{xx}-H_{yy}}\right].
\end{equation}
We estimated an angle $\theta_{ij}$ for each of the $m$\,$\times$\,$n$ pixels in a velocity channel map, where the indices $i$ and $j$ run over the pixels along the $x$- and $y$-axis, respectively.
This angle, however, is only meaningful in regions of the map that are rated as filamentary according to selection criteria based on the values of $\lambda_{-}$ and on the noise properties of the data \citep[see, for example,][]{schisano2020}.

We conducted the Hessian analysis in $10\deg$\,$\times$\,$10\deg$ tiles and across velocity channels.
We selected the filamentary structures in each tile based on the criterion $\lambda_{-}$\,$<$\,$0$.
Additionally, we selected regions where $I(l,b,v)$\,$>$\,3$\sigma_{I}$, where $\sigma_{I}$ corresponds to the RMS noise presented in Sect.~\ref{section:observations}.
Following the method introduced in \cite{planck2014-XXXII}, we further selected the filamentary structures depending on the values of the eigenvalue $\lambda_{-}$ in noise-dominated data portions.
We estimated $\lambda_{-}$ in five velocity channels with low signal-to-noise ratios and determined the minimum value of $\lambda_{-}$ in each.
We used the median of these five $\lambda_{-}$ values as the threshold value, $\lambda^{C}_{-}$.
We employed the median to reduce the effect of outliers. 
Still, in general, the values of $\lambda_{-}$ in the noise-dominated channels are similar, and this selection does not imply any loss of generality.
We exclusively considered regions of each velocity channel map where $\lambda_{-}$\,$<$\,$\lambda^{C}_{-}$, which corresponds to the selection of filamentary structures with curvatures in $I(l,b,v)$ larger than those present in the noise-dominated channels.

Once the filamentary structures were selected, we used the angles derived from Eq.~\ref{eq:theta} to study their orientation with respect to the Galactic plane.
For a systematic evaluation of the preferential orientation, we applied the projected Rayleigh statistic ($V$) \citep[see, for example,][]{batschelet1981}, which is a test to determine whether the distribution of angles is nonuniform and peaked at a particular angle.
This test is equivalent to the modified Rayleigh test for uniformity proposed by \cite{durandANDgreenwood1958} for the specific directions of interest $\theta$\,$=$\,$0\deg$ and $90\deg$ \citep{jow2018}, such that $V$\,$>$\,0 or $V$\,$<$\,0 correspond to preferential orientations parallel or perpendicular to the Galactic plane, respectively.
It is defined as
\begin{equation}\label{eq:myprs}
V = \frac{\sum^{n,m}_{ij}w_{ij}\cos(2\theta_{ij})}{\sqrt{\sum^{n,m}_{ij}w_{ij}/2}},
\end{equation}
where the indices $i$ and $j$ run over the pixel locations in the two spatial dimensions $(l,b)$ for a given velocity channel and $w_{ij}$ is the statistical weight of each angle $\theta_{ij}$.

The values of $V$ lead to significance only if sufficient clustering is found around the orientations $\theta$\,$=$\,$0\deg$, and 90$\deg$.
The null hypothesis implied in $V$ is that the angle distribution is uniform or centered on a different orientation angle.
In the particular case of independent and uniformly distributed angles, and for a large number of samples, values of $V$\,$\approx$\,$1.64$ and $2.57$ correspond to the rejection of the null hypothesis with a probability of 5\% and 0.5\%, respectively \citep{batschelet1972}.
A value of $V$\,$\approx$\,2.87 is roughly equivalent to a 3$\sigma$ confidence interval.
We present our analysis results in terms of the mean orientation angle, $\left<\theta\right>$, and the Rayleigh statistic, $Z$, in App.~\ref{appendix:method}.

In our application, we accounted for the spatial correlations introduced by the telescope beam by choosing $w_{ij}$\,$=$\,$(\Delta x/D)^{2}$, where $\Delta x$ is the pixel size and $D$ is the diameter of the derivative kernel chosen to calculate the gradients.
This selection guarantees that $V$ is independent of the map pixelization. 
We note, however, that the correlation across scales in the ISM makes it very difficult to estimate the absolute statistical significance of $V$.

\section{Results}\label{section:results}

\subsection{Prevalent CO filament orientation}\label{section:COorientation}

We present the global trends for the \CO\ line emission anisotropy throughout the Galactic disk in the right-hand panel of Fig.~\ref{fig:lvdiagrams1}.
It is apparent that most of the tiles show $V$\,$>$\,0.
It is also noticeable that the highest-$V$ tiles are clustered for positive LOS velocities in QI and negative toward the fourth Galactic quadrant (QIV), which corresponds to the inner Milky Way.

\begin{figure}[ht]
\centerline{\includegraphics[width=0.5\textwidth,angle=0,origin=c]{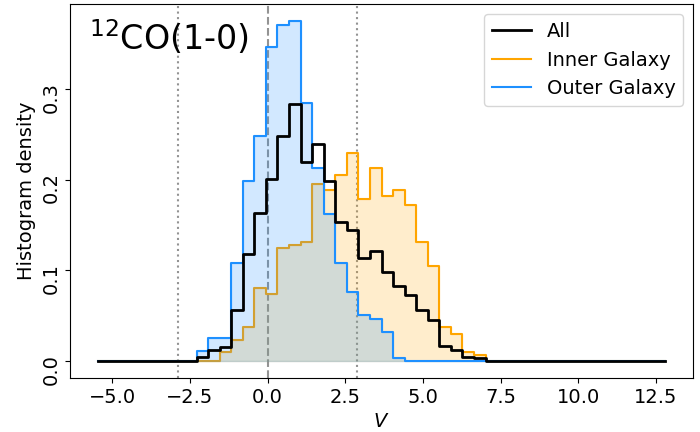}}
\centerline{\includegraphics[width=0.5\textwidth,angle=0,origin=c]{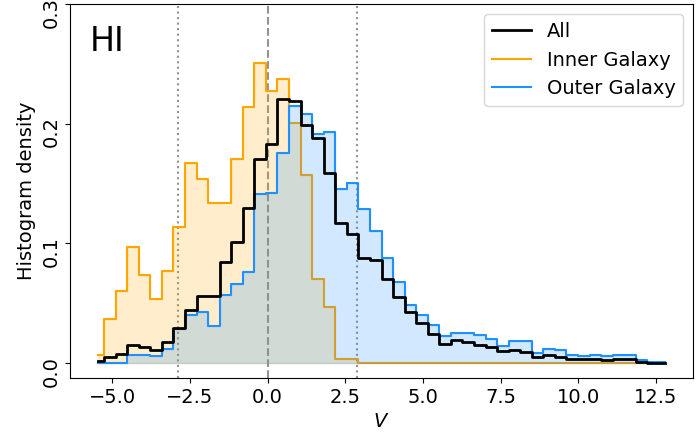}}
\caption{Normalized histogram of the $V$ distribution across 10\deg\,$\times$\,10\deg\,$\times$\,1.29-\kps\ tiles for \CO\ ({\it top}), corresponding to the values presented on the left-hand panel of Fig.~\ref{fig:lvdiagrams1}, and H{\sc i}, corresponding to figure~A.3 in \cite{soler2022}.}
\label{fig:histPRS}
\end{figure}

Figure~\ref{fig:histPRS} shows the distribution of $V$ across tiles.
Roughly 85\% show $V$\,$>$\,0.
Around 20\% of the tiles show $V$\,$>$\,2.87, which is the 3$\sigma$ equivalent for a preferential orientation parallel to the Galactic plane.
None of the tiles shows $V$\,$<$\,$-2.87$, and less than 0.26\% of the tiles shows $V$\,$<$\,$-1.64$, thus indicating that a preferential orientation perpendicular to the Galactic plane is rare in the CO emission.
These results are very much alike for different initial signal-to-noise ratio selections and other Galactic plane segmentations, for similar kernel sizes, as shown in App.~\ref{appendix:method}.

Figure~\ref{fig:histPRS} also confirms that most tiles with significant preferential orientation parallel to the Galactic plane are found toward the inner Galaxy. 
Toward the outer Galaxy, there is no evident global anisotropy parallel or perpendicular to the Galactic plane, as quantified by the predominant $|V|$\,$<$\,2.87 and the mean orientation angles reported in App.~\ref{appendix:method}.
The latter outcome is in agreement with the results reported in \cite{dib2009}, which presents the analysis of positions angles (PAs) of MCs in the 50\arcsec-resolution Five College Radio Astronomy Observatory (FCRAO) \CO\ survey of the Outer Galaxy \citep{heyer1998,heyer2001} and found that their orientations are random.

\begin{figure}[ht]
\centerline{\includegraphics[width=0.5\textwidth,angle=0,origin=c]{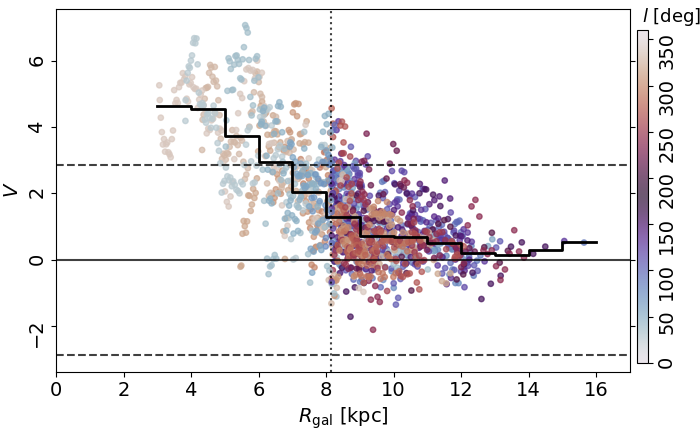}}
\caption{Distribution of $V$ as a function of Galactocentric distance ($R_{\rm gal}$).
Each marker corresponds to a 10\deg\,$\times$\,10\deg\,$\times$\,1.29\,\kps-tile on the left-hand panel of Fig.~\ref{fig:lvdiagrams1}.
The colors represent each tile's central Galactic longitude ($l$).
The black solid line shows the $V$ mean value in 1-kpc-wide $R_{\rm gal}$ bins.
The dashed vertical line indicates the radius of the Solar orbit, $R_{\rm \odot}$\,$=$\,8.15\,kpc.}
\label{fig:PRSvsRgal}
\end{figure}

Figure~\ref{fig:PRSvsRgal} shows the variations in $V$ with Galactocentric distance ($R_{\rm gal}$), computed using the central $v_{\rm LSR}$ of each tile and assuming circular motions around the Galactic center following the Galactic rotation model presented in \cite{reid2019}.
Following \cite{soler2022} and other preceding works, 
we excluded from the distance reconstruction the Galactic longitude ranges 
$l$\,$<$\,15\deg, $l$\,$>$\,345\deg, and 165\deg\,$<$\,$l$\,$<$\,195\deg.
This selection aims to minimize the $R_{\rm gal}$ uncertainties produced by non-circular motions in the direction of the Galactic center and anti-center \citep{hunter2024}, as well as excluding the sphere of influence of the Galactic bar \citep{goller2025}.

The scatter plot in Fig.~\ref{fig:PRSvsRgal} further illustrates that orientations preferentially parallel to the Galactic plane in CO emission are mainly found at $R_{\rm gal}$\,$<$\,$R_{\rm \odot}$.
The asymmetry in the $V$ distribution around $R_{\rm gal}$\,$=$\,$R_{\rm \odot}$ indicates that Heliocentric distance, $d$, is not systematically producing lower levels, as a consequence of the fixed angular size of the derivative.
If the filamentary structures used to quantify the anisotropy were randomized with increasing $d$ by the spatial filtering of 
the derivative kernel, as illustrated with the fBm models in App.~\ref{app:fBM}, one would expect lower $V$ at opposite sides of  $R_{\rm gal}$\,$=$\,$R_{\rm \odot}$ in Fig.~\ref{fig:PRSvsRgal}, 
Instead, we find that the position to the Galactic center is more relevant in the level of anisotropy quantified by $V$.

The progressive decrease in the $V$ mean values with increasing $R_{\rm gal}$ in Fig.~\ref{fig:PRSvsRgal} can be interpreted as a decrease in the anisotropy level in the \CO\ distribution with increasing distance from the Galactic center.
The prevalence of $|V|$\,$<$\,2.87 for $R_{\rm gal}$\,$\gtrsim$\,9\,kpc indicates the predominance of random orientations toward the outer Galaxy. 
The Galactocentric radius at which the mean anisotropy levels fall below the significance threshold $V$\,$=$\,2.87 is around 6\,kpc.
However, there are a few tiles with $V$\,$>$\,2.87 at larger $R_{\rm gal}$ and up to roughly 10\,kpc from the Galactic center.

Visual inspection of the $l$-$v_{\rm LSR}$ distribution of $V$ in Fig.~\ref{fig:lvdiagrams1} suggests that the high-$V$ tiles are coincident with some of the positions along the tracks corresponding to the Scutum-Centaurus (Sct-Cen) spiral arm toward QI and QIV.
However, high-$V$ tiles are also found outside of the spiral arm tracks.
Tiles along the Perseus (Per) arm show low $V$ values, suggesting that the orientation of CO filaments is random for the emission along that $l$-$v_{\rm LSR}$ track.

The local structures in CO emission, understood as either those at $|v_{\rm LSR}$|\,$\lesssim$\,20\,\kps\ in Fig.~\ref{fig:lvdiagrams1} or $R_{\rm gal}$\,$\approx$\,$R_{\odot}$ in Fig.~\ref{fig:psrSNRtest}, do not show a preferential orientation.
The signal-to-noise ratio in the \cite{dame2001} observations is insufficient to identify enough significant emission toward the Outer, Outer Scutum Centaurus (OSC), and extended outer spiral arms, where the most prominent H{\sc i} filaments parallel to the Galactic plane were identified in \cite{soler2022}.
However, these portions of the Galactic plane have been sampled with higher sensitivity in other higher angular resolution surveys, as we discuss in the following section.

\begin{figure*}[ht]
\centerline{
\includegraphics[width=0.99\textwidth,angle=0,origin=c]{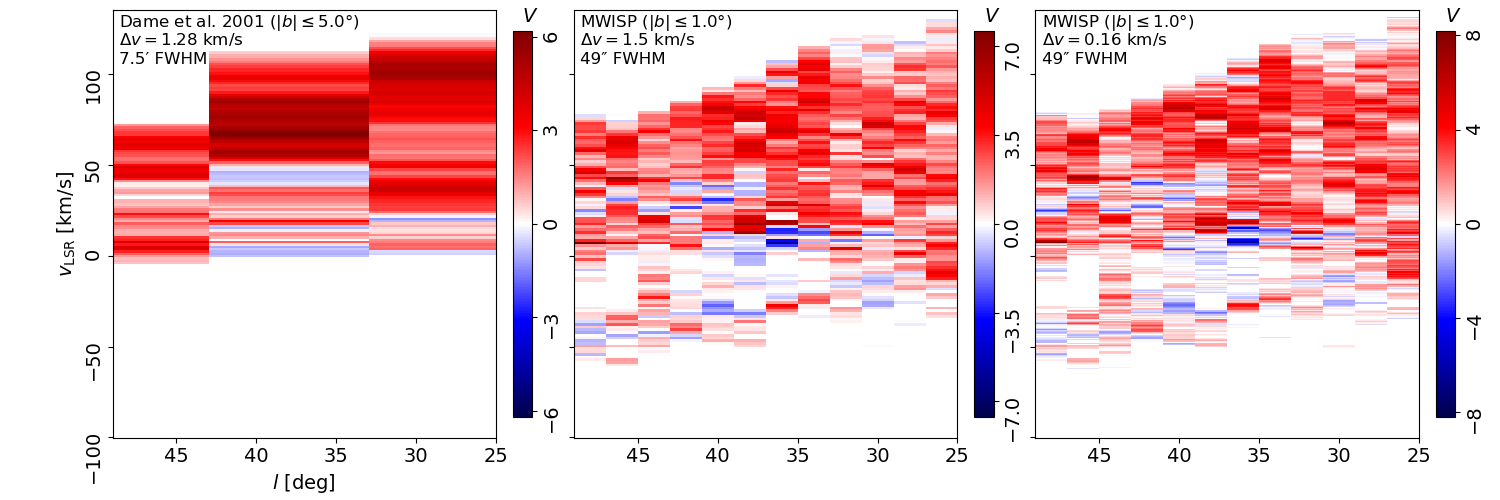}
}
\caption{Comparison of the filament anisotropy identified in the 10\deg\,$\times$\,10\deg\ regions for the 7\parcm5-resolution \CO\ observations in the \cite{dame2001} ({\it left}), and the 2\deg\,$\times$\,2\deg\ regions for the 49\arcsec-resolution \CO\ observations in the MWISP survey for velocity channel widths 1.5\,\kps\ ({\it middle}) and 0.16\,\kps\ ({\it right}), as reported in \cite{soler2021}.}
\label{fig:lvdiagramsMWISPsoler2021}
\end{figure*}

\subsection{Comparison to higher-resolution observations}

To determine the effects of the 7\parcm5 FWHM resolution in the \cite{dame2001} composite survey in the results of our anisotropy analysis, we considered the portion of the Galactic plane covered by the MWISP survey \CO\ observations at roughly nine times higher angular resolution, although it is limited to the range $|b|$\,$<$\,1\deg.
The MWISP data set was previously analyzed in \cite{soler2021}, but we reproduced their analysis for comparison with the global trends across the plane.

Figure~\ref{fig:lvdiagramsMWISPsoler2021} presents the preferential orientation throughout the $^{12}$CO MWISP observations split in 2\deg\,$\times$\,2\deg\ regions and 1.49-\kps-wide tiles.
As previously reported in \cite{soler2021}, the majority of the tiles at $v_{\rm LSR}$\,$>$\,0\,\kps\ show $V$\,$>$\,2.87, suggesting that the emission is preferentially oriented parallel to the plane toward that portion of the first Galactic quadrant.
Toward the outer Galaxy, $v_{\rm LSR}$\,$<$\,0\,\kps, there is no predominant preferential orientation.

\begin{figure*}[ht]
\centerline{\includegraphics[width=0.9\textwidth,angle=0,origin=c]{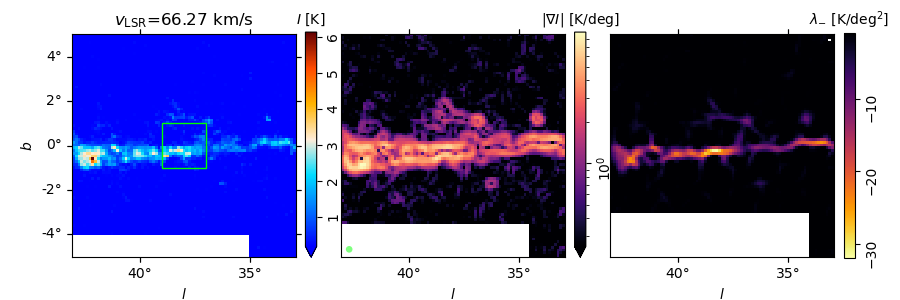}}
\centerline{\includegraphics[width=0.9\textwidth,angle=0,origin=c]{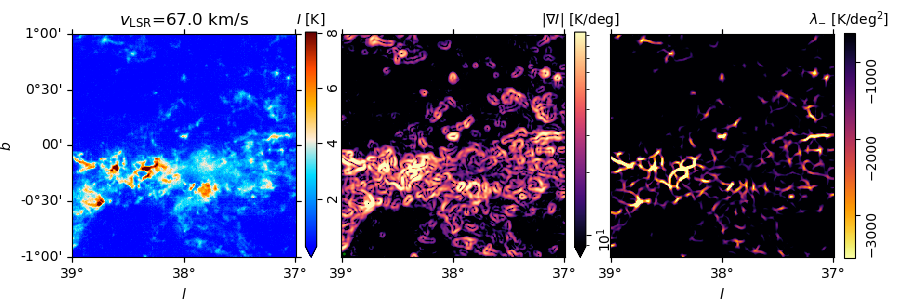}}
\caption{Example of a tile with CO emission structures mostly parallel to the Galactic plane, as identified by the highest $V$ values on the left panel of Fig.~\ref{fig:lvdiagrams1}.
The top panels show the \CO\ line emission at 7\parcm5 FWHM resolution in the \cite{dame2001} composite survey.
The bottom panels show the \CO\ line emission at 55\arcsec\ FWHM resolution in the MWISP survey toward the region marked with the green square in the top panel.
The left, middle, and right panels show the line emission map, its gradient, and the filamentary structures identified by the second eigenvalue of the Hessian matrix, Eq.~\eqref{eq:lambda}, respectively.
}
\label{fig:maxVexample}
\end{figure*}

The coincidence in the preferential filament orientations for $v_{\rm LSR}$\,$>$\,0\,\kps\ in the \cite{dame2001} and MWISP observations suggest that the global anisotropy in the emission distribution is the same across one order of magnitude in spatial scales.
Figure~\ref{fig:maxVexample} illustrates the persistence of the global orientation of the structures in the two surveys.
Although the MWISP data reveals clouds and features with different orientations in the CO emission at higher resolution, their preferential orientation maintains the global orientation pattern identified at larger spatial scales in the \cite{dame2001} data.
The structures identified with the Hessian matrix method for the two surveys, shown on the right panels of Fig.~\ref{fig:maxVexample}, confirm that this technique is not simply identifying the same large-scale structure in both surveys but revealing small-scale structures that have the same anisotropy as the coarse-resolution data.

\begin{figure*}[ht]
\centerline{\includegraphics[width=0.9\textwidth,angle=0,origin=c]{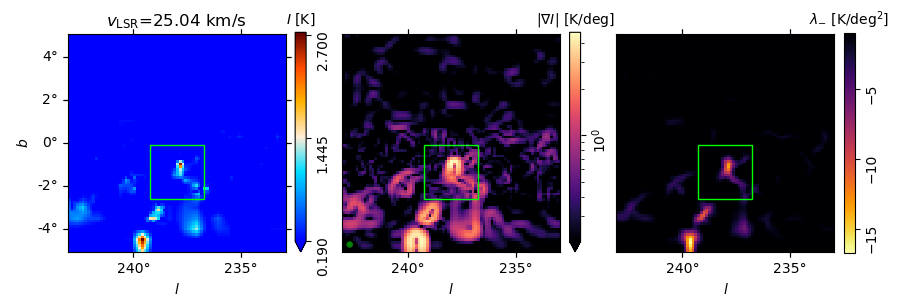}}
\centerline{\includegraphics[width=0.9\textwidth,angle=0,origin=c]{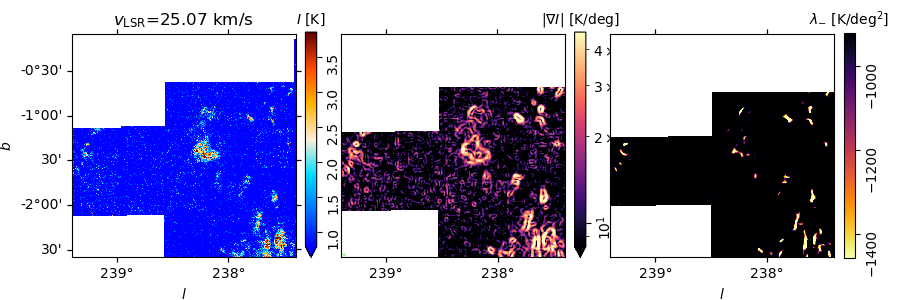}}
\caption{Same as Fig.~\ref{fig:maxVexample}, but for the tile with the most prevalent CO structures perpendicular to the Galactic plane, characterized by the lowest $V$ values on the left panel of Fig.~\ref{fig:lvdiagrams1}.
The bottom panels show a zoom-in with the 55\arcsec-resolution \CO\ line observations in the FQS \citep{benedettini2020}.
}
\label{fig:minVexample}
\end{figure*}

We also compared low- and high-resolution $^{12}$CO emission data for a tile with preferentially vertical structures, as identified by the lowest $V$ in the $lv$-diagram on the right panel of Fig.~\ref{fig:lvdiagrams1}.
This tile, shown in Fig.~\ref{fig:minVexample}, is covered by the FQS, which reveals that the coarse structures identified in the \cite{dame2001} CO data resolve into smaller structures that preserve the large-scale orientation.
This region shows filamentary structures perpendicular to the Galactic plane instead of the prevailing alignment with the disk.
This location coincides with the position of GSH\,238+00+09, a nearby major superbubble toward Galactic longitude around 238\deg\ \citep{heiles1998}, where \CO\ emission is faint and dust extinction is singularly low \citep[see, for example,][]{soler2025}.

\subsection{Comparison with the H{\sc i} filament orientation}\label{section:HIcomparison}

\begin{figure}[ht]
\centerline{\includegraphics[width=0.5\textwidth,angle=0,origin=c]{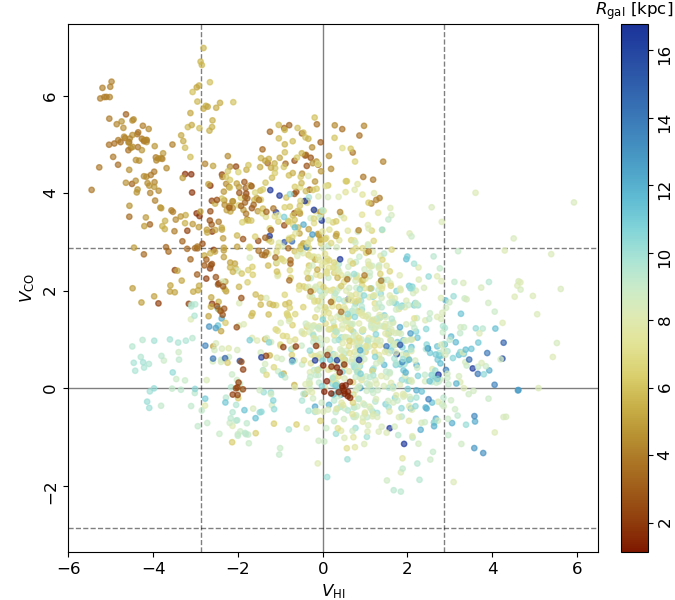}}
\caption{Comparison of the preferential orientation obtained with the projected Rayleigh statistic ($V$, Eq.~\ref{eq:myprs}) applied to the \CO\ and H{\sc i} observations.
The latter corresponds to the \cite{soler2020} results for the same tile size and derivative kernel size.
The colors indicate the Galactocentric distances of each tile.
}
\label{fig:HIandCOprs2}
\end{figure}

Figure~\ref{fig:HIandCOprs2} presents the comparison between the CO filament orientation trends reported on the right panel of Fig.~\ref{fig:lvdiagrams1} and the equivalent analysis of filamentary structure in the H{\sc i} emission, presented in the middle of figure~A.3 in \cite{soler2022}.
The scatter plot shows that the values of $V_{\rm CO}$\,$>$\,2.87 tend to appear in tiles with $V_{\rm HI}$\,$<$\,$-2.87$, that is, the most evident anisotropy parallel to the Galactic plane in the CO emission is found where the H{\sc i} is preferentially vertical.
The tile extension in Galactic latitude can explain this trend; within a $|b|$\,$<$\,5\deg\ regions, there are both relatively high heights above the disk, which are dominated by H{\sc i} emission in the vertical structures identified in \cite{soler2022}, and low scale heights, where CO emission is more prevalent.
Yet, the comparison of the preferential orientation in the MWISP data and the 40\arcsec\ H{\sc i} observations the HI/OH/Recombination line (THOR) survey of the inner Milky \citep{beuther2016,wang2020hi} toward the Galactic latitude band $|b|$\,$<$\,1\deg\ also indicate a dissimilar behavior of the two tracers at smaller scales and lower latitude \citep{soler2020,soler2021}.
The latter result suggests that the difference in the anisotropy in the structures traced by H{\sc i} and CO emission is not exclusively due to a scale-height selection effect.

\section{Discussion}\label{section:discussion}

\subsection{CO orientations toward the inner and outer Galaxy}

One of the main findings in our study of the anisotropy traced by the orientation of filamentary structure in \CO\ emission is the dissimilar behavior in the inner and outer Milky Way.
It is illustrated in the predominantly positive $V$ around $v_{\rm LSR}$\,$>$\,0\,\kps\ toward QI and $v_{\rm LSR}$\,$<$\,0\,\kps\ toward QIV in the right-hand panel of Fig.~\ref{fig:lvdiagrams1} and the low $V$ for other regions and velocity ranges.
It is also noticeable in the $V$ trends presented in Fig.~\ref{fig:histPRS} and Fig.~\ref{fig:PRSvsRgal}.

The extend of our results toward the outer Galaxy is limited by the \cite{dame2001} data sensitivity.
However, the higher-sensitivity and angular resolution observations in MWISP confirm the random orientations of CO structures beyond the Solar circle, as illustrated in Fig.~\ref{fig:lvdiagramsMWISPsoler2021}.
The analysis presented in \cite{dib2009} also suggests that this may be the case for outer Galaxy CO structures in the second Galactic quadrant (QII).

The portion of the Galactic plane sampled with higher resolution and sensitivity with MWISP data also shows predominantly horizontal CO structure in the emission toward the inner Galaxy, as also illustrated in Fig.~\ref{fig:lvdiagramsMWISPsoler2021}. 
This means that the structures parallel to the Galactic plane at 7\parcm5 resolution do not resolve into randomly oriented structures at 49\arcs\ resolution, and the prevalent anisotropy is similar across roughly one decade of spatial scales.
This is an expected behavior given the correlations across scales introduced by turbulence in the ISM \citep[see, for example,][]{elmegreenANDscalo2004,hennebelleANDfalgarone2012}.

What we denote as inner and outer Galaxy corresponds to two domains in the CO distribution. 
The CO mass in the outer galaxy is between 2.3 and 2.7 times lower than in the inner Galaxy \citep[see][and references therein]{heyerANDdame2015}.
The CO mass surface density profile also drops by a factor of roughly five and more after $R_{\rm gal}$\,$\approx$\,7\,kpc, as illustrated in figure~7 in \cite{heyerANDdame2015}.
The CO scale height also increases from roughly a full-width half-maximum (FWHM) around 100\,pc for $R_{\rm gal}$\,$\lesssim$\,$R_{\odot}$ to a few times that value beyond the Solar circle, as shown in the top panel of figure~6 in \cite{heyerANDdame2015}.
The midplane height is also displaced by more than 100\,pc for  $R_{\rm gal}$\,$\gtrsim$\,$R_{\odot}$, as presented in the bottom panel of figure~6 in \cite{heyerANDdame2015}.

The difference in the radial extent of CO can be attributed to variations in the mid-plane pressure, as recently shown in the numerical simulation presented in \cite{smith2023}.
That means that the structures traced by CO in the outer Galaxy are less constrained to remain in the plane than structures within the Solar circle. 
Neutral atomic hydrogen (H{\sc i}) observations indicate velocity dispersions between 12 and 15\,\kps to toward the central parts of nearby face-on spiral galaxies and between 4 and 6\,\kps\ toward their outskirts \citep[e.g.,][]{ianjamasimanana2015}.
Turbulent velocity fluctuations associated with this observed vertical velocity field in the outer Galaxy can potentially be responsible for randomizing CO structures' orientation in regions of lower vertical pressure.

\cite{soler2022} pointed to the energy and momentum input from clustered SNe explosions as the source of the prevalent vertical (perpendicular to the plane) H{\sc i} structures toward the inner Galaxy.
Toward the outer Galaxy, where the population of SN-precursor high-mass stars is lower, the H{\sc i} filamentary structures are predominantly parallel to the Galactic plane.
In the case of CO, the combination of the lower coupling of the dense gas with the input from SNe and the higher mid-plane pressure is the most plausible source of CO anisotropy parallel to the Galactic plane in the inner Galaxy.
Toward the outer Galaxy, other sources of interstellar turbulence, such as accretion from the galactic halo and global galactic motions \citep[see, for example][]{klessen2010,meidt2018}, can produce motions that randomize the orientation of the dense gas traced by CO with respect to the Galactic plane.

\subsection{Relation to GMFs and other Galactic filamentary structures}

The preferential orientation parallel to the Galactic plane for CO emission can also be related to the potential effect of spiral density waves \citep{linANDshu1964}. 
Toward the inner Galaxy, where the spiral density waves are presumed to be stronger due to the higher densities of gas and stars, the flows would be actively redirected along the arm.
This provides a compressive component that allows preexisting small, molecular clouds to coagulate and build up mass to form larger MCs and trigger instabilities in the atomic gas that ultimately lead to the formation of more MCs \citep[see, for example,][]{roberts1969,pringle2001}. 
The structures assembled through this mechanism would be long and preferentially aligned with the plane, as the ``Nessie'' filament in the Milky Way and other elongated molecular structures in nearby galaxies \citep[see, for example,][]{goodman2014,meidt2023}.
Yet, numerical studies show that a spiral potential is not indispensable to produce filamentary structures aligned with the disk; the galaxies' differential rotation stretches out any density enhancement that extends out in the radial direction 
\citep[see, for example,][]{smith2014,duarte-cabral2017}.

The extension of our analysis to identify individual structures, as spiral arms, is complicated by the overlap of multiple structures in position-position distance (PPD) into a single velocity channel in position-position velocity (PPV), which is more prevalent towards those regions than towards the inner Galaxy.
It is plausible that the overlap of multiple PPD structures in PPV is producing the marked anisotropy with respect to the Galactic plane we observed toward the inner Galaxy.
Such an anisotropy in PPV channels would not be observed if it was not initially present in the corresponding portions of the PPP volume.
However, linking an individual PPP structure, such as a spiral arm, to the overall PPV anisotropy reported in Fig.~\ref{fig:lvdiagrams1} is not straightforward and is beyond this work's scope.

\cite{zucker2018} presented the physical properties of large-scale Galactic filaments identified in mid- and far-infrared emission and in emission from CO and other high-density tracers.
Most of the objects in that sample display low inclinations with respect to the Galactic plane, which agrees with our results for the general anisotropy in the CO emission.

\cite{colombo2021} identified and characterized the filamentary structures in the $^{12}$CO(2--1) emission observations in the Outer Galaxy High-Resolution Survey (OGHReS) and reported that large-scale filaments in the outer Galaxy show on average masses and linear masses around one order of magnitude lower than similar structures toward the inner Galaxy.
This behavior has been previously reported in the comparison of MCs in the inner and outer Galaxy, and has been attributed to variations in metallicity or dust-to-gas ratio with $R_{\rm gal}$ \citep[see, for example,][]{ladaANDdame2020}.
Our results suggest that in addition to these effects, global physical conditions may vary with increasing distance from the Galactic center, as evidenced by different anisotropy found in H{\sc i} and CO LOS velocity channels corresponding to the inner and outer Milky Way.

\subsection{Comparison to numerical simulations}

Direct comparison of the present analysis with synthetic observations is not yet available, primarily due to the limited spatial resolution of the galactic-scale simulations required to reproduce simultaneously the density structures sampled by CO and encompass large-scale dynamics.
However, numerical models of 1-kpc-side boxes have been used to study the gas properties produced by galactic winds and fountains from a star-forming, stratified ISM.
For example, \cite{kimANDostriker2018} shows the differential effect of SN feedback in different gas phases, with warm and cold ISM clouds entrained by a high-velocity, low-density hot wind, maintaining the cold gas at lower scale heights than the warmer and hot components.
The numerical experiments in a stratified box presented in \cite{girichidis2018b} indicate that magnetic fields increase the disk scale height by the factor of a few and delay the formation of dense and molecular gas.
Although their analysis does not explicitly address the orientation of the structures traced by CO, included as part of the chemical network in their model, it is apparent from the column density projections that dense gas structures remain close to the mid-plane and are only momentarily parallel to the Galactic before their orientation is randomized by the effect of SN explosions.

The Galactic-scale simulations of multiphase ISM presented in \cite{smith2020} find that spiral-arm-like structures and differential rotation preferentially align filamentary structures, while strong feedback randomizes the orientations. 
Our observations, however, indicate that most of the CO filamentary structures aligned with the plane are found toward the inner Galaxy, where SN feedback is more concentrated.
The analysis of synthetic observations of large-scale Galactic filaments presented in \cite{zucker2019} indicates that most of the coherent and elongated structures display low inclinations with respect to the Galactic plane.

As expected from the relatively low SFR in the Milky Way, we find no prevalent signature of Galactic outflows in the molecular gas. 
This is due not only to the lack of input from direct SN feedback but also to cosmic ray (CR) pressure.
Recent numerical experiments, such as those presented in \cite{armillotta2024}, indicate that CR pressure near the midplane is comparable to other pressure components in the gas, but the scale height of CRs is far larger and can efficiently accelerate warm gas above and below the plane.
More recently, \cite{kjellgren2025} show that CRs can drive weak but sustained outflows throughout a Galactic disk simulation. 
Such an effect would not be evident in the scale height but can lift the gas in the entire disk and, through subsequent fountain flows, favor gas mixing and produce colder extraplanar gas \citep[see, for example,][]{fraternali2006}.

Figure~\ref{fig:HIandCOprs2} shows that, in general, the orientation of the CO filaments does not correspond to that of the H{\sc i} structures characterized in \cite{soler2022}.
This dissimilar behavior can be related to the distinct swept-up mass and total momentum for gas at different densities \citep[see, for example,][]{kimANDostriker2015,martizzi2015}.
Both the H{\sc i} and CO are subject to the same large-scale gravitational potential, establishing the Galactic plane as the central axis of symmetry. 
However, H{\sc i} is less dense and can couple better to the vertical momentum input from stellar feedback, producing the chimneys we see in the vertical direction and leaving the densest CO clouds closer to the plane, as seen, for example, in the multiphase ISM numerical simulations presented in \cite{walch2015} and 
\cite{girichidis2016}\footnote{Videos available at the \href{https://girichidis.com/index.php/research-overview/silcc}{SILCC website}}.


Our observations do not imply, however, that all molecular gas is following the asymmetry implied by that identified in the CO emission. 
Most likely, there is also molecular gas in the H{\sc i} structures perpendicular to the Galactic plane.
We simply cannot see it in CO because of the low column densities leading to the destruction of the CO molecule, whereas molecular hydrogen (H$_{2}$) and dust may remain intact. 
Recent James Webb Space Telescope (JWST) observations of polycyclic aromatic hydrocarbons (PAH) emission reveal dust filamentary structures extending to few-kiloparsec altitudes above the disks of edge-on galaxies NGC 891 and M82 \citep[see, for example,][]{chastenet2024,fisher2025}.
The presence of dust at these high altitudes  challenges the current understanding of the transport mechanisms involved, as it suggests that small dust grains survive for several tens of million years after having been ejected by galactic winds in the disk-halo interface.
An analysis of the relation of between PAH features and CO clouds in M82 indicates that it is possible that CO emission is not tracing the full budget of molecular gas in that galaxy, perhaps as a consequence of photoionization and/or emission suppression of CO molecules due to hard radiation fields from the central starburst \citep{villanueva2025}.

Identifying the molecular gas not traced by CO in the vertical structures traced by H{\sc i} toward the inner Galaxy is highly challenging.
Yet, our study indicates that the Milky Way is unlikely to display the prominent chimneys traced by CO seen in other galaxies, such as NGC253 \citep{krieger2019}.
Our results also highlight the anisotropy in the CO emission and its changes with Galactocentric radius, which is harder to disentangle in extragalactic but can provide additional information about the transport mechanisms in Galactic winds.

\section{Conclusions}\label{section:conclusions}

We presented a study of the \CO\ line emission anisotropy across the Milky Way's disk to examine the effect of stellar feedback and galactic dynamics on the distribution of the dense interstellar medium. 
We found that the structures sampled with this tracer are predominantly parallel to the Galactic plane in the inner Galaxy, in clear contrast with the primarily perpendicular orientation of the structures traced by neutral atomic hydrogen (H{\sc i}) emission toward the same regions.
Our results suggest that Galactic molecular winds, traced by elongated CO structures perpendicular to the disks of nearby galaxies, are not currently prevalent in the Milky Way.
Extending our analysis to portions of the Galactic plane sampled at higher resolution with other surveys in narrower latitude bands, we found that the structures traced by CO emission are still preferentially oriented parallel to the Galactic plane in the inner Galaxy, but randomly oriented beyond the Solar circle.

One hypothesis for the observed behavior is that it results from the lower mid-plane pressure in the outer Galaxy and the dissimilar effect of SN feedback on the diffuse and dense gas.
In that scenario, diffuse structures traced by H{\sc i} are efficiently expelled off the disk by clustered supernova while the denser gas traced by CO tends to remain close to the midplane in regions of high mid-plane pressure like the inner Galaxy.
The mid-plane pressure is lower toward the outer Galaxy, as evidenced by the decreasing CO surface density. 
Thus, the structures traced by CO are less restrained by the disk anisotropy and show random orientations, most likely caused by the turbulent fluctuations introduced by processes different to the SN feedback prevalent in the inner galaxy. 

Another possibility is that we are measuring an effect related to the CO as a gas tracer. 
In the inner Galaxy, we may not see the molecular gas in the ISM lifted off the midplane because CO not a good tracer of that gas phase.
In the outer Galaxy, it is also possible that CO is not tracing a more extended molecular component, as suggested by the OH emission observations of the outer Milky Way \citep[see, for example,][]{busch2021}.

In contrast with studies based on segmenting CO emission to define objects such as clouds and spiral arms, our analysis focused on a general emission property to determine variations throughout the Galaxy. 
In addition to the variations of the CO surface density, mid-plane height, and scale height with the Galactocentric radius, the anisotropy emission also changes with increasing distance from the Galactic center.
Whether this behavior corresponds to a variation in the physical mechanisms acting on the gas or a change in the CO emission distribution calls for additional studies, particularly of galactic-scale numerical simulations.
However, our results already show that the position with respect to the galactic center influences the main molecular gas tracer, elucidating a potential link between the conditions leading to star formation and the large-scale galactic environment.  

\begin{acknowledgements}
The European Research Council funds JDS via the ERC Synergy Grant ``ECOGAL -- Understanding our Galactic ecosystem: From the disk of the Milky Way to the formation sites of stars and planets'' (project ID 855130, PIs P.~Hennebelle, R.~S.~Klessen, S.~Molinari, L.~Testi).
Some of the crucial discussions that led to this work took place under the program Milky-Way-Gaia of the PSI2 project, which is funded by the IDEX Paris-Saclay, ANR-11-IDEX-0003-02.
JDS thanks the following people for their encouragement and conversation: Fabian Walter, Henrik Beuther, Bob Benjamin, Philipp Girichidis, Karin Kjellgren, Lilly Kormann, and Michelangelo Pantaleoni.
RSK also acknowledges financial support from the German Excellence Strategy via the Heidelberg Cluster ``STRUCTURES'' (EXC 2181 - 390900948) as well as from from the German Ministry for Economic Affairs and Climate Action in project ``MAINN'' (funding ID 50OO2206).  RSK  thanks the 2024/25 Class of Radcliffe Fellows for the stimulating discussions. 
The computations for this work were performed at the Max-Planck Institute for Astronomy (MPIA) {\tt astro-node} servers.
{\it Software}: {\tt astropy} \citep{astropy2018}, {\tt TurbuStat} \citep{koch2019turbustat}, {\tt SciPy} \citep{2020SciPy-NMeth}, {\tt magnetar} \citep{magnetar2020}.

\end{acknowledgements}

\bibliographystyle{aa}
\bibliography{COfilaments.bbl}

\clearpage
\appendix

\section{The Hessian analysis method}\label{appendix:method}

In the main body of this paper, we presented an analysis of the anisotropy traced by the filamentary structures selected using the Hessian analysis method.
In this section, we illustrate the generality of our conclusions against the Hessian method input parameters and the circular statistic used to report the anisotropy results.

\subsection{Data selection}\label{appendix:data}

\begin{figure}[ht]
\centerline{\includegraphics[width=0.495\textwidth,angle=0,origin=c]{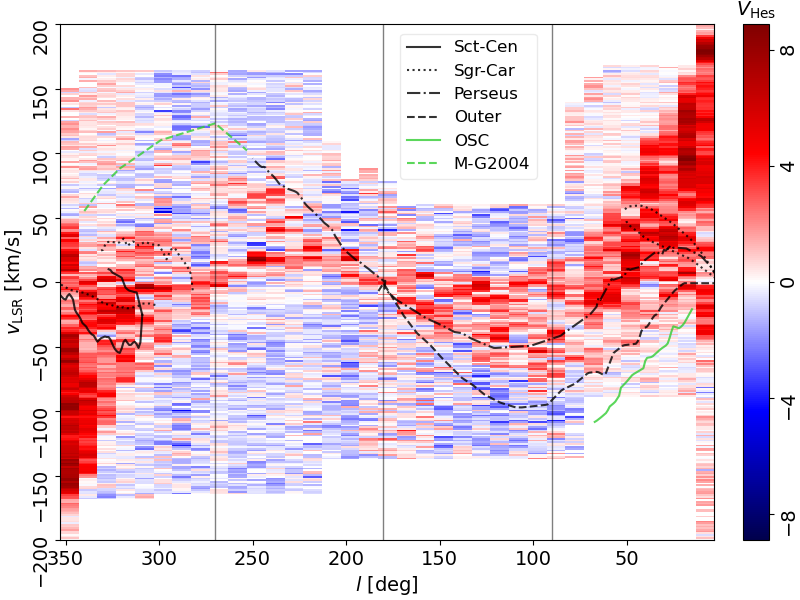}}
\centerline{\includegraphics[width=0.495\textwidth,angle=0,origin=c]{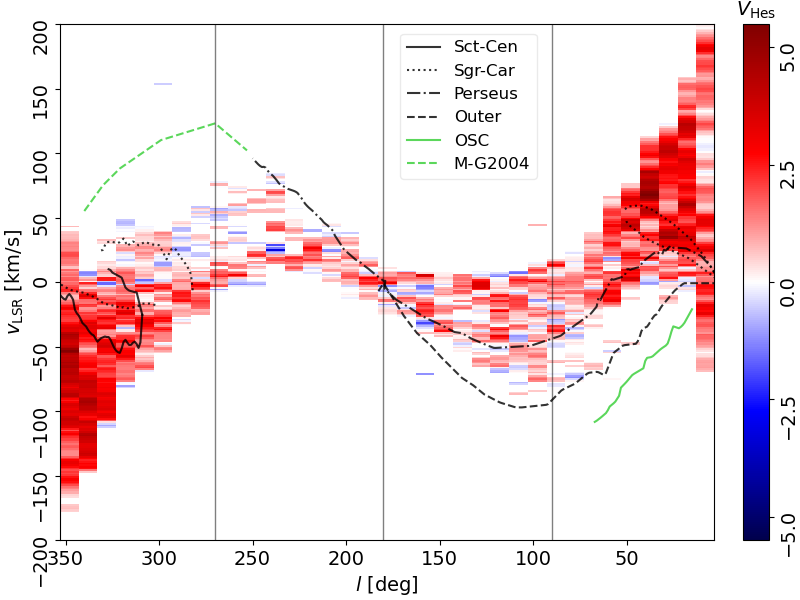}}
\caption{Same as the right-hand panel of Fig.~\ref{fig:lvdiagrams1}, but without Monte Carlo realizations and excluding regions with signal-to-noise ratio below 1.0 ({\it top}) and 3.0 ({\it bottom}).}
\label{fig:psrSNRtest}
\end{figure}

We propagated the uncertainties in the \CO\ line measurements by using Monte Carlo realizations of the input data, as described in Sec.~\ref{section:methods}.
An alternative error handling is obtained by applying signal-to-noise cuts to the input data.
Figure~\ref{fig:psrSNRtest} shows the results of the anisotropy analysis reported on the right-hand panel of Fig.~\ref{fig:lvdiagrams1} without Monte Carlo realizations but with an initial cut by the signal-to-noise ratio ($I/\sigma_{I}$) in the input data.
The results obtained excluding pixels with $I/\sigma_{I}$\,$<$\,1.0, presented in the top panel of Fig.~\ref{fig:psrSNRtest}, show a spurious signal in $V$ for low-intensity channels. 
A more stringent cut of pixels with $I/\sigma_{I}$\,$<$\,3.0, shown in the bottom panel of Fig.~\ref{fig:psrSNRtest}, results in a clearer contrast in $V$ between regions with significant signal and noise-dominated channels. 
Comparison between the bottom panel of Fig.~\ref{fig:psrSNRtest} and the right-hand panel of Fig.~\ref{fig:lvdiagrams1} indicates that the Monte Carlo sampling averages out some of the signals in low-\CO\ emission channels.
For example, for $v_{\rm LSR}$\,$<$\,0\,\kps\ beyond the Perseus arm and toward the third Galactic quadrant (QIII), where the \CO\ emission is faint, probably due to the expulsion of material by GSH\,238+00+09.

The results reported on the right-hand panel of Fig.~\ref{fig:lvdiagrams1} correspond to an arbitrary segmentation of the Galactic employed to evaluate the anisotropy.
We considered alternative segmentations considering two separate options. 
First, offsetting in Galactic longitude the location of the 10\deg\,$\times$\,10\deg\ tiles employed in the main body of the paper. 
Second, employing 5\deg\,$\times$\,5\deg\ centered on $b$\,$=$\,0\deg.

\begin{figure}[ht]
\centerline{\includegraphics[width=0.495\textwidth,angle=0,origin=c]{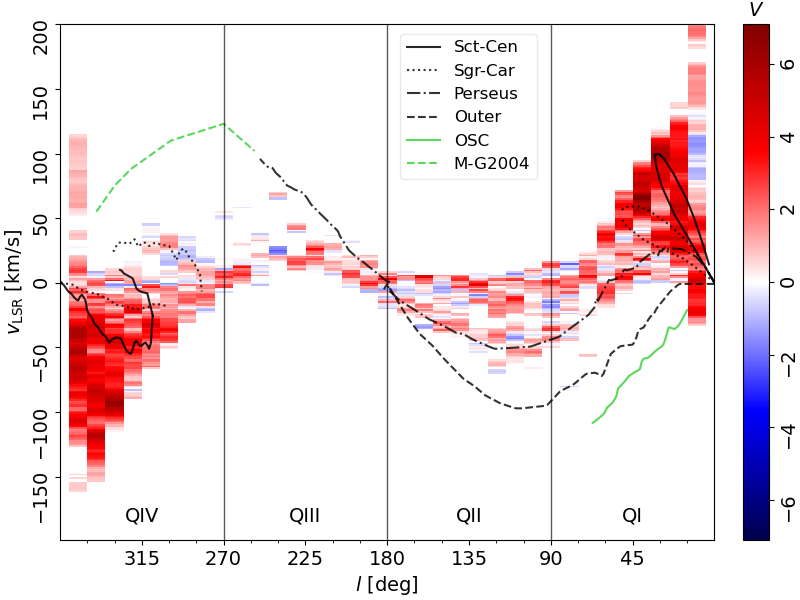}}
\caption{Same as the right-hand panel of Fig.~\ref{fig:lvdiagrams1}, but a segmentation into 10\deg\,$\times$\,10\deg\ tiles offset by 2\pdeg5.}
\label{fig:lvprs10by10offset}
\end{figure}

Figure~\ref{fig:lvprs10by10offset} presents the results obtained by adding an offset of 2\pdeg5 to the central position of the 10\deg\,$\times$\,10\deg\ tiles used in Fig.~\ref{fig:lvdiagrams1}.
Some of regions show different $V$, most likely as a result of splitting anisotropic structures in a velocity channel and combining them with less anisotropic structures in the same tile.
However, the prevalence of high $V$ toward the inner Galaxy and the random orientation of structures beyond the Solar circle is independent of the central position of the 10\deg\,$\times$\,10\deg\ tiles, as illustrated for the particular selection presented in Fig.~\ref{fig:lvprs10by10offset} and others tested in our analysis.

\begin{figure}[ht]
\centerline{\includegraphics[width=0.495\textwidth,angle=0,origin=c]{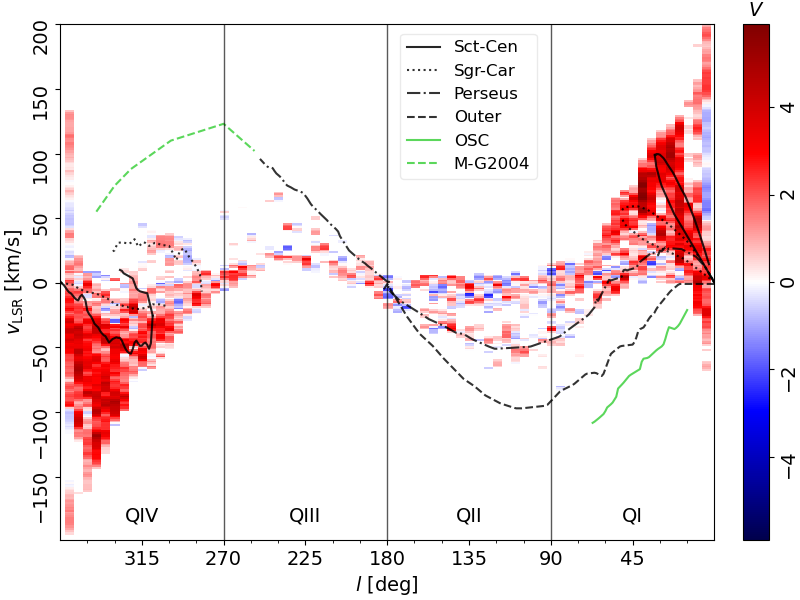}}
\caption{Same as the right-hand panel of Fig.~\ref{fig:lvdiagrams1}, but 5\deg\,$\times$\,5\deg\ tiles centered on $b$\,$=$\,0\deg.}
\label{fig:lvprs5by5}
\end{figure}

The main reason for using the 10\deg\,$\times$\,10\deg\ was the direct comparison with the results of the H{\sc i} emission analysis presented in \citep{soler2022}.
However, the 7\pdeg5 angular resolution of the \cite{dame2001} composite survey allows for a finer segmentation.
Figure~\ref{fig:lvprs5by5} presents the results obtained using a segmentation of the Galactic plane into 5\deg\,$\times$\,5\deg\ tiles centered on $b$\,$=$\,0\deg.
It is evident by visual comparison with the right-hand panel of Fig.~\ref{fig:lvdiagrams1} that the trends discussed in the main body of the paper are also present in the region $|b|$\,$\leq$\,5\deg\ and are persistent in the finer $l$ segmentation.

Some features in Fig.~\ref{fig:lvprs5by5} may inspire additional interpretations. 
For example, the prevalence for $V$\,$>$\,0 in a significant segment of the Perseus arm.
However, this behavior is not seen in the MWISP data for the same spiral arm track in the $lv$-diagram, as illustrated in Fig.~\ref{fig:lvdiagramsMWISPsoler2021}.

\subsection{Hessian method parameter selection}

Two parameters are used to identify filamentary structures using the Hessian matrix.
First, the derivative kernel size, which set the spatial scale at which the filaments are singled out.
Second is the critical threshold of the eigenvalue $\lambda_{-}$, which defines the minimum curvature of the features diagnosed as filaments.

\begin{figure}[ht]
\centerline{\includegraphics[width=0.495\textwidth,angle=0,origin=c]{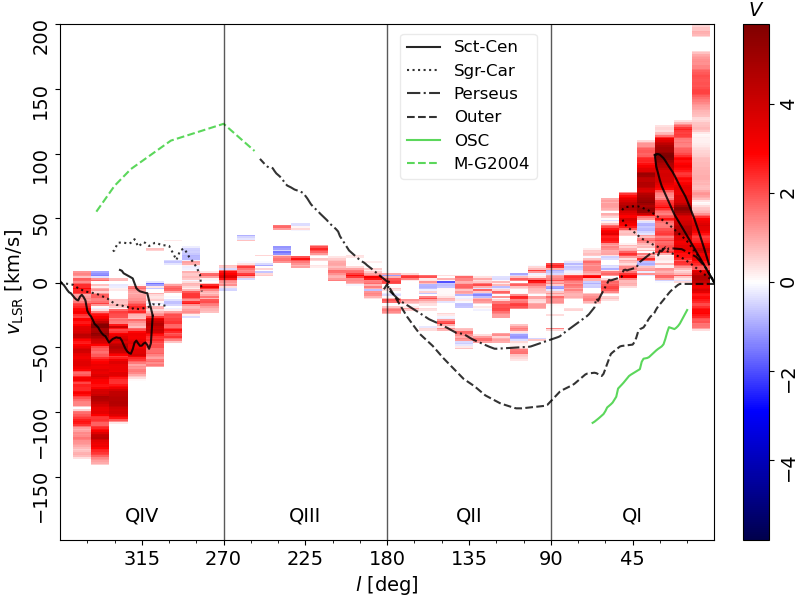}}
\caption{Same as the right-hand panel of Fig.~\ref{fig:lvdiagrams1}, but 30\arcmin\ FWHM derivative kernel.}
\label{fig:lvprs_ksz30arcmin}
\end{figure}

The resolution of the observations sets the minimum derivative kernel size.
The analysis of the MWISP and FQS, discussed in Sec.~\ref{section:results}, shed light on the results obtained for smaller kernel sizes. 
However, the areas of the sky covered by these observations are smaller than those in the \cite{dame2001} composite survey.
We show an example of the results obtained with a larger kernel size in Fig.~\ref{fig:lvprs_ksz30arcmin}, where we employed a 30\arcmin\ FWHM derivative kernel.
The amplitude of $V$ in this example is lower, as expected from the lower number of independent gradient vectors for the same tile area. 
However, the main features in the $V$ $lv$-diagram are coincident with those reported in the right-hand panel of Fig.~\ref{fig:lvdiagrams1}.
These results are similar for larger kernel sizes up to angular scales of roughly 1\pdeg25, for which almost all the tiles show values below the significance level $|V|$\,$\approx$\,2.87.

We identified a critical curvature value $\lambda^{\rm c}_{-}$\,$\approx$\,$-1.0$\deg\ using the median of the values obtained in the Hessian analysis of noise-dominated velocity channels.
This value may overestimate the noise level in some areas, given the sensitivity variation across the \citep{dame2001} composite survey.
However, it provides a common denominator to filter out potential spurious filamentary structures introduced by the noise.

\subsection{Statistics for reporting preferential orientations}

\begin{figure}[ht]
\centerline{\includegraphics[width=0.495\textwidth,angle=0,origin=c]{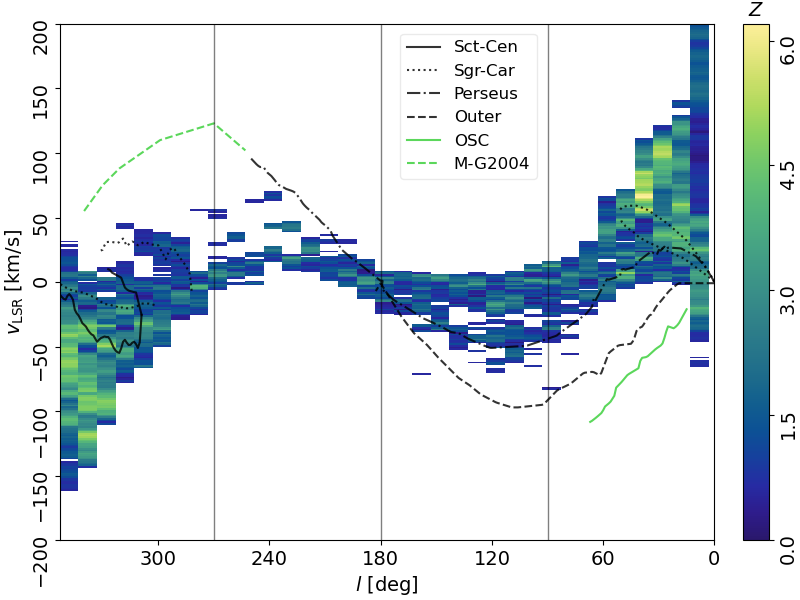}}
\centerline{\includegraphics[width=0.495\textwidth,angle=0,origin=c]{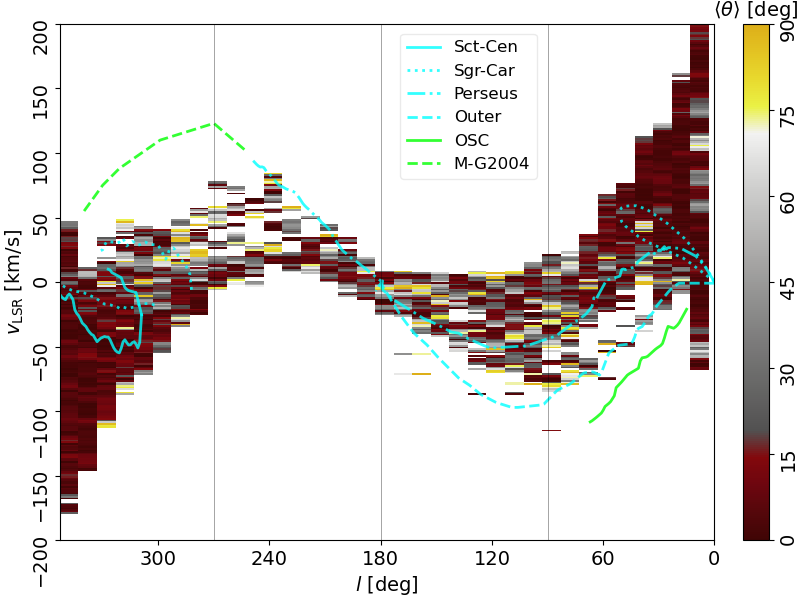}}
\caption{Same as the right-hand panel of Fig.~\ref{fig:lvdiagrams1}, but for the Rayleigh statistic ($Z$) and the mean orientation angle ($\left<\theta\right>$).}
\label{fig:lvdiagramsRSandTheta}
\end{figure}

On the right-hand panel of Fig.~\ref{fig:lvdiagramsRSandTheta}, we reported the anisotropy in terms of the projected Rayleigh statistic, Eq.~\eqref{eq:myprs}, an optimal estimator for determining the clustering of the angle data around a particular orientation \citep[see, for example,][]{batschelet1972,mardia2009directional,jow2018}.
In this appendix, we present the results in terms of two additional angular quantities: the Rayleigh statistic, $Z$, and the mean orientation angle, $\left<\theta\right>$, shown in Fig.~\ref{fig:lvdiagramsRSandTheta}.

\begin{figure}[ht]
\centerline{\includegraphics[width=0.495\textwidth,angle=0,origin=c]{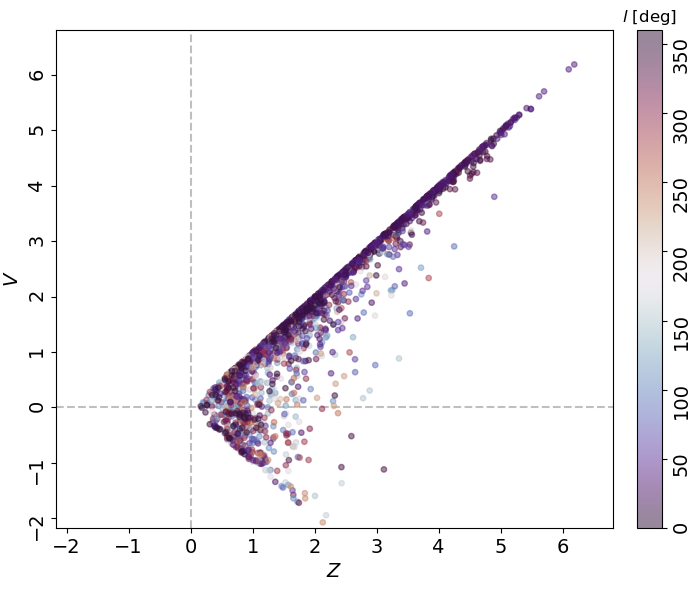}}
\caption{Relation between $V$ and $Z$ of the tiles presented in the right-hand panel of Fig.~\ref{fig:lvdiagrams1} and the top panel of Fig.~\ref{fig:lvdiagramsRSandTheta}.
The colors indicate the central Galactic longitude of each tile.}
\label{fig:scatterRSandPRS}
\end{figure}

The Rayleigh statistic \citep[or Rayleigh test;][]{batschelet1972} is defined as 
\begin{equation}\label{eq:myrs}
Z \equiv \sqrt{V^{2}+W^{2}}, 
\end{equation}
where $V$ correspond to the projected Rayleigh statistic defined in Eq.~\eqref{eq:myprs} and 
\begin{equation}\label{eq:w}
W \equiv \frac{\sum^{n,m}_{ij}w_{ij}\sin(2\theta_{ij})}{\sqrt{\sum^{n,m}_{ij}w_{ij}/2}}.
\end{equation}
This statistic represents the general case of the projected Rayleigh statistic without setting a preferential orientation to test against.
That means that high values of $Z$ correspond to clustered data around a particular angle. 
Comparison between the $V$ and $Z$ for our anisotropy analysis, presented in Fig.~\ref{fig:scatterRSandPRS}, show that the high $Z$ corresponds to high $V$, which is another wait of showing that the filament orientation angle clustering is around 0\deg.
This can also be confirmed by the correspondence in the distributions of $V$ and $Z$ presented in the right-hand panel of Fig.~\ref{fig:lvdiagrams1} and the top panel Fig.~\ref{fig:lvdiagramsRSandTheta}.

The clustering in filament orientation around 0\deg\ is also shown by the mean orientation angle, which is defined as
\begin{equation}\label{eq:meantheta}
\left<\theta\right> \equiv 0.5\arctan\left(\frac{W}{V}\right). 
\end{equation}
The distribution of $\left<\theta\right>$ in the $lv$-diagram presented on the bottom panel of Fig.~\ref{fig:lvdiagramsRSandTheta} is an alternative way of visualizing our anisotropy analysis results.
However, that visualization does not consider that $\left<\theta\right>$ can be ill-defined in tiles with a homogeneous distribution of angles or with few significant orientation measurements, as illustrated in App.~\ref{app:fBM}. 

\section{Gradients instead of filaments as a measure of anisotropy}\label{app:gradients}

In the main body of the paper, we reported the results obtained using the filament orientation as a measure of anisotropy.
In this appendix, we present the results of the anisotropy analysis based on the preferential orientation of intensity gradients.
By definition, the gradient vectors indicate the direction perpendicular to the contours in a scalar field, such as the line emission intensity. 
Thus, one can quantify the anisotropy in the distribution of the structures in that scalar field by evaluating the clustering of the orientations of the gradient vectors rotated by 90\deg. 
\cite{micelotta2021} used numerical simulations to show that the characterization of scalar fields using gradients or filaments carries complementary information about the orientation of structures.
However, in our application, it is unclear {\it a priori} that the two methods for anisotropy evaluation lead to the same results.

We computed the intensity gradients in each 10\deg\,$\times$\,10\deg\,$\times$\,1.29\,\kps\ channel using the same derivative kernel employed in the computation of the partial derivatives in the Hessian matrix, as described in Sec.~\ref{section:methods}.
We then rotated the gradient orientation angles by 90\deg\ and used the resulting values as an input in Eq.~\ref{eq:myrs} to compute $V$, which we label as $V_{\rm Grad}$ to distinguish it from the Hessian analysis results.
Error propagation is made using Monte Carlo sampling of the noise, as in the Hessian matrix analysis.

\begin{figure}[ht]
\centerline{\includegraphics[width=0.495\textwidth,angle=0,origin=c]{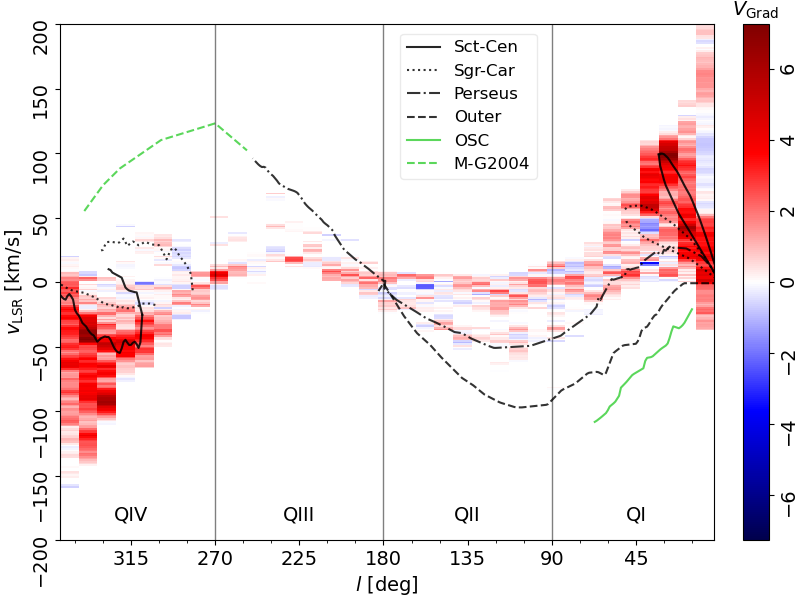}}
\caption{Same as the right-hand panel of Fig.~\ref{fig:lvdiagrams1}, but for the \CO\ emission orientation derived from the intensity gradients.
}
\label{fig:lvprsGradients}
\end{figure}

Figure~\ref{fig:lvprsGradients} shows the anisotropy results across \CO\ line emission channels evaluated using the gradients.
The distribution of $V_{\rm Grad}$ is similar to the results reported in Fig.~\ref{fig:lvdiagrams1}.
The main difference is that $V_{\rm Grad}$ is even closer to zero for tiles toward the outer Galaxy, highlighting the random orientation of the structures traced by \CO\ in that portion of the Milky Way.

\section{Toy model tests}\label{app:fBM}

\begin{figure*}[ht]
\centerline{\includegraphics[width=0.95\textwidth,angle=0,origin=c]{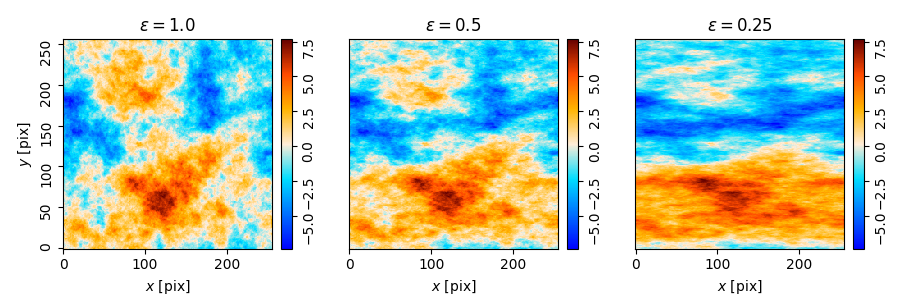}}
\caption{Synthetic emission maps produced using fBm realizations with a power spectrum slope $\alpha$\,$=$\,3.0, anisotropy angle $\theta$\,$=$\,0\deg, and ellipticities $\epsilon$\,$=$\,1.0, 0.5, and 0.25, shown from left to right.}
\label{fig:fBMs}
\end{figure*}

We evaluated the effects of the beam size and overlapping of structures by employing fractional Brownian Motion (fBm) simulations of a 2D scalar field \citep{bruntANDheyer2002,miville2008}.
In practice, the fBm generates a random field that follows a power-law spectrum with slope $-\alpha$ and anisotropy defined by ellipticity $\epsilon$ and the angle $\theta$.
For this test, we chose a Kolmogorov-like spectrum, $\alpha$\,$=$\,3, and tested ellipticities that range between $\epsilon$\,$=$\,1.0, which corresponds to a completely isotropic scalar field, and $\epsilon$\,$=$\,0.75, 0.5, and 0.25.
An example of the relationship between $\epsilon$ and the scalar field morphology is presented in Fig.~\ref{fig:fBMs}, where it is evident that the realizations with the same initial random but different values of $\epsilon$ display anisotropy with respect to the preferential orientation $\theta$\,$=$\,0\deg.

\begin{figure}[ht]
\centerline{\includegraphics[width=0.49\textwidth,angle=0,origin=c]{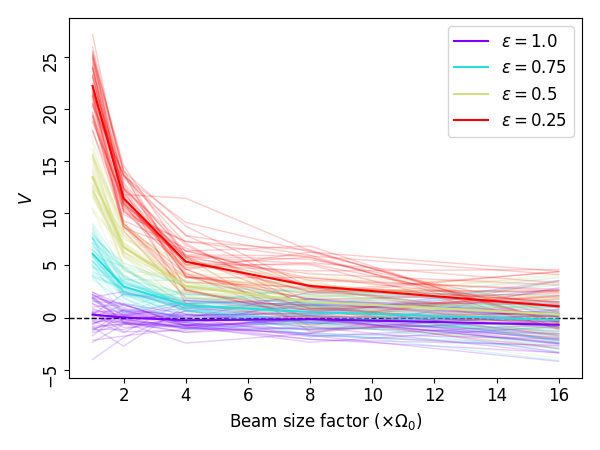}}
\centerline{\includegraphics[width=0.49\textwidth,angle=0,origin=c]{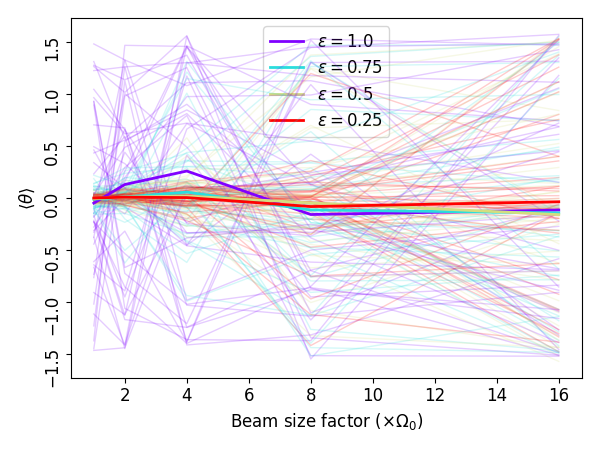}}
\caption{Anisotropy quantified by the projected Rayleigh statistic ($V$, {\it top}) and the mean orientation angle ($\left<\theta\right>$, {\it bottom}) for the filamentary structures in fBm maps with ellipticities, $\epsilon$, and for different beam sizes above the derivative kernel size $\Omega_{0}$\,$=$\,18\arcmin.
For each color, the narrow lines correspond to the 48 fBm realizations and the broad line to the mean value of the realizations.}
\label{fig:fBMsSummary}
\end{figure}

We produced 48 fBm scalar fields with 81 pixels $\times$ 81 pixels, which matches the size of the tiles analyzed in the main body of the paper.
Each realization corresponds to a different initial random seed.
We analyzed these fields using the Hessian method and employing different kernel sizes, expressed as multiples of the initial kernel size $\Omega_{0}$\,$=$\,18\arcm.
The results presented in Fig.~\ref{fig:fBMsSummary} indicate that for the same initial derivative kernel width, the anisotropy quantified with $V$ roughly escalates linearly with $\epsilon$.
For larger kernel sizes, which can be interpreted as placing the maps at further distances or selecting larger angular scales for the filament selection, the values of $V$ progressively decrease as the number of independent gradient vectors in each map is reduced. 

Figure~\ref{fig:fBMsSummary} also shows the analysis of the fBm realizations in terms of $\left<\theta\right>$, Eq.~\eqref{eq:meantheta}.
The large fluctuations in $\left<\theta\right>$ for $\epsilon$ illustrate the limitations of this metric to identify anisotropy.
This is further demonstrated by the similarity of the mean values of $\left<\theta\right>$ for realizations with different ellipticities. 


\raggedright

\end{document}